\documentclass{pasj00}
\begin{document}
\SetRunningHead{Yamasaki, et al. }{X-ray halo around NGC~4631}
\Received{2008/07/30}%{yyyy/mm/dd}
\Accepted{2008/08/29}%{yyyy/mm/dd}

\title{X-ray Halo Around the Spiral Galaxy NGC~4631 Observed with Suzaku}

%%% begin:list of authors
% Do NOT capitalize all letters in "textsc".
\author{
Noriko Y. \textsc{Yamasaki} \altaffilmark{1},
Kosuke \textsc{Sato}\altaffilmark{2},
Ikuyuki {\sc Mitsuishi}\altaffilmark{1},
Takaya {\sc Ohashi}\altaffilmark{3}
}%

%% \thanks{Example: Present Address is xxxxxxxxxx}}
\altaffiltext{1}{Institute of Space and Astronautical Science, Japan Aerospace Exploration Agency (ISAS/JAXA),\\
3-1-1 Yoshinodai, Sagamihara, Kanagawa, 229-8510}
\email{yamasaki@astro.isas.jaxa.jp}
\altaffiltext{2}{Graduate School of Natural Science and Technology, Kanazawa University, Kakuma, Kanazawa, Ishikawa, 920-1192}
\altaffiltext{3}{Department of Physics, Tokyo Metropolitan University, \\
1-1 Minami-Osawa, Hachioji, Tokyo, 192-0397}
%%% end:list of authors

%% `\KeyWords{}' always has to be placed before `\maketitle'.
 %Do NOT move this preamble from here!

\maketitle

\begin{abstract}
Suzaku observation of the edge-on spiral galaxy NGC 4631 confirmed its
X-ray halo extending out to about 10 kpc from the galactic disk. The
XIS spectra yielded the temperature and metal abundance for the disk
and the halo regions. 
The observed abundance pattern for O, Ne, Mg, Si and Fe
is consistent with the metal yield from type II supernovae, with an O mass
of about $10^{6} M_\odot$ contained in the halo. These features imply
that metal-rich gas produced by type II supernova is brought into the
halo region very effectively, most likely through a galactic
wind. Temperature and metal abundance may be affected by charge
exchange and dust. An upper limit for the hard X-ray flux was
obtained, corresponding to a magnetic field higher than $0.5 \mu$G\@.
\end{abstract}

\section{Introduction}

For the understanding of the chemical evolution of galaxies and
clusters of galaxies, precise knowledge about the metal production
from Type Ia (SN Ia) and Type II (SN II) supernovae (SNe) is of vital
importance. X-ray imaging spectroscopy of supernova remnants (SNRs),
galaxies and intra-cluster medium  (ICM) has shown that metal abundances
generally vary from source to source. However, \citet{sato07a}
determined the abundances of O, Ne, Mg, Si, and Fe in several galaxy
clusters based on  Suzaku observations, and found  that the
abundance patterns are commonly represented by a combination of type
Ia and II supernovae  products with an occurrence number ratio of
$1:3.5$, based on the theoretical yields from SN II and SN Ia. In
order to further look into the past history of the two types of
supernova, we need more precise knowledge about the metal yields from
the different SNe types. It is, however, quite difficult to
extract pure supernova products, in particular for the X-ray emitting
hot gas. When we observe young supernova remnants (SNR), there is
always a mixture of supernova ejecta and surrounding ISM and line
intensities also strongly depend on the ionization condition. One way
to provide a good constraint is to observe X-ray halos around
starburst galaxies, which are considered to be maintained by the
enhanced SNe II activity in the recent time.

Recent X-ray observations of starforming galaxies showed that metals
are contained in the extended hot halo of the galaxies. For M82 and
NGC 253, RGS spectra for several sliced regions along the outflow axis
showed   lines from highly ionized O, Ne, Mg, Si and Fe
\citep{read02,bauer07}. The observed intensity ratios of the lines
indicates that the gas is  cooling as it travels outward from the
galaxy disk and that the gas around NGC 253 could be partly out of
ionization equilibrium. Suzaku observations of a ``cap'' region
of M82, which is 11.6 kpc north of the galaxy and is possibly a
termination region of the hot-gas outflow, showed a spectrum
consisting of emission lines from O through Fe \citep{tsuru07}. These
spectral features strongly suggest that fresh metal-rich gas produced
in the starforming region is flowing out mainly along the minor axis
of galaxies. However, metal abundances in the halo gas has so far not
been well-constrained. RGS data are  limited in statistics, and both the EPIC
and the ACIS instruments do  not have sufficient energy resolution below 1
keV\@ (e.g.\ \cite{tuellmann06}). Suzaku offers a good
opportunity for measuring the metallicity of  the outflowing hot gas.

NGC~4631 is a nearby Sc/SBd galaxy with  an edge-on morphology with the
distance estimated to be about 7.5 Mpc, where $1'$ corresponds to
2.2~kpc. Estimated mass by the Tully-Fisher relation is $2.6\times
10^{10} M_{\odot}$ \citep{strickland04}. The inclination and
position angle are $81^\circ$ and $356^\circ$ respectively. This
galaxy is suitable for Suzaku observations of  the X-ray halo
maintained by its SN activity. With its radio halo \citep{hummel90}
and warm IR ratio, it is classified as a mild disk-wide starburst
galaxy \citep{golla94b}.
An extended X-ray halo was  discovered by  ROSAT  \citep{wang95},
and it has been well studied with  Chandra
\citep{wang01,oshima03,strickland04} and XMM-Newton
\citep{tuellmann06}. The size of the halo is several arcmin  and
no X-ray counterpart for the central AGN has been  detected
\citep{strickland04}. The association of an H$\alpha$ filament with the
X-ray emission has been   discovered \citep{wang01}, and FUSE also
detected  O\emissiontype{VI} lines from a region at $2'$ above the disk
\citep{otte03}. These strongly suggest that a halo around NGC 4631 is
the site of galactic outflow or fountain, where the gas is floating up
from the disk by the SN energy input and possibly cooling.

Another important feature of the halo is the extended synchrotron
radio emission, which is  observed from several star-forming
galaxies (e.g.\ \cite{veilleux05}). These radio halos are populated
with relativistic electrons together with magnetic fields of an order
of $10 \mu$G\@. The radial orientation of the magnetic field lines
suggests that the field has been carried into the halo region by the
outflow of hot gas \citep{golla94a}. In some cases, the hot gas may be
confined in the halo when the magnetic pressure exceeds the thermal
pressure  \citep{wang95}. To constrain how the non-thermal energy is distributed
in the halo, it is important to look into the possibility of  non-thermal
emission in the hard X-ray band.

Throughout this paper we adopt the Galactic hydrogen column density of
$N_{\rm H} = 1.27\times10^{20}$ cm$^{-2}$ \citep{dickey90} in the
direction of NGC~4631\@. The solar abundance table is given by
\citet{anders89}, and the errors are the 90\% confidence limits for
a single interesting parameter.

\section{Observation and Data Reduction}

\subsection{Observation}

\begin{figure*}
\begin{minipage}{0.45\textwidth}
%%\FigureFile(\textwidth,\textwidth){n4631_0.6_0.7.eps}
\FigureFile(\textwidth,\textwidth){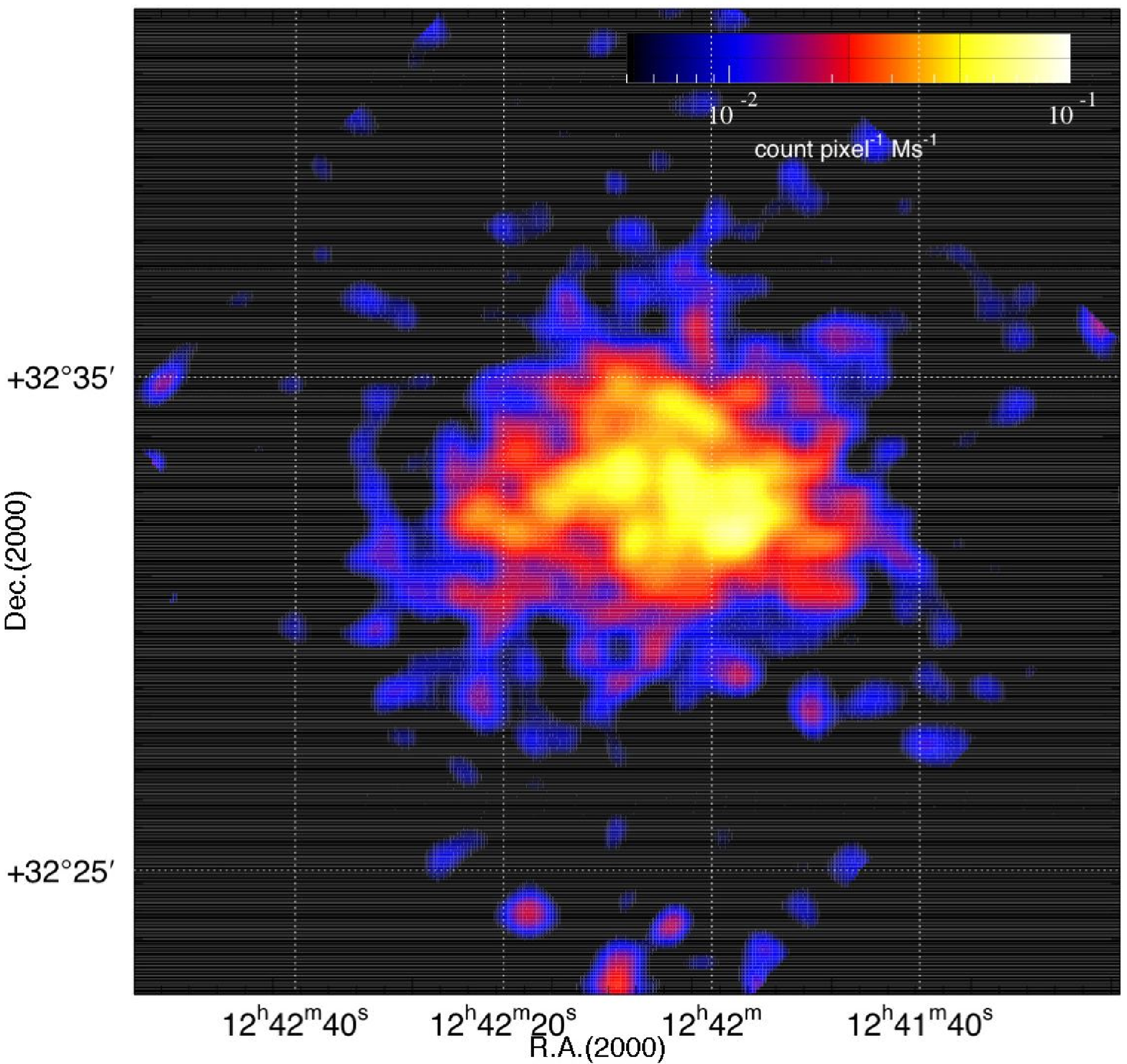}
\end{minipage}
\hfill
\begin{minipage}{0.45\textwidth}
%%\FigureFile(\textwidth,\textwidth){n4631_0.5_2_opt_cont.eps}
\FigureFile(\textwidth,\textwidth){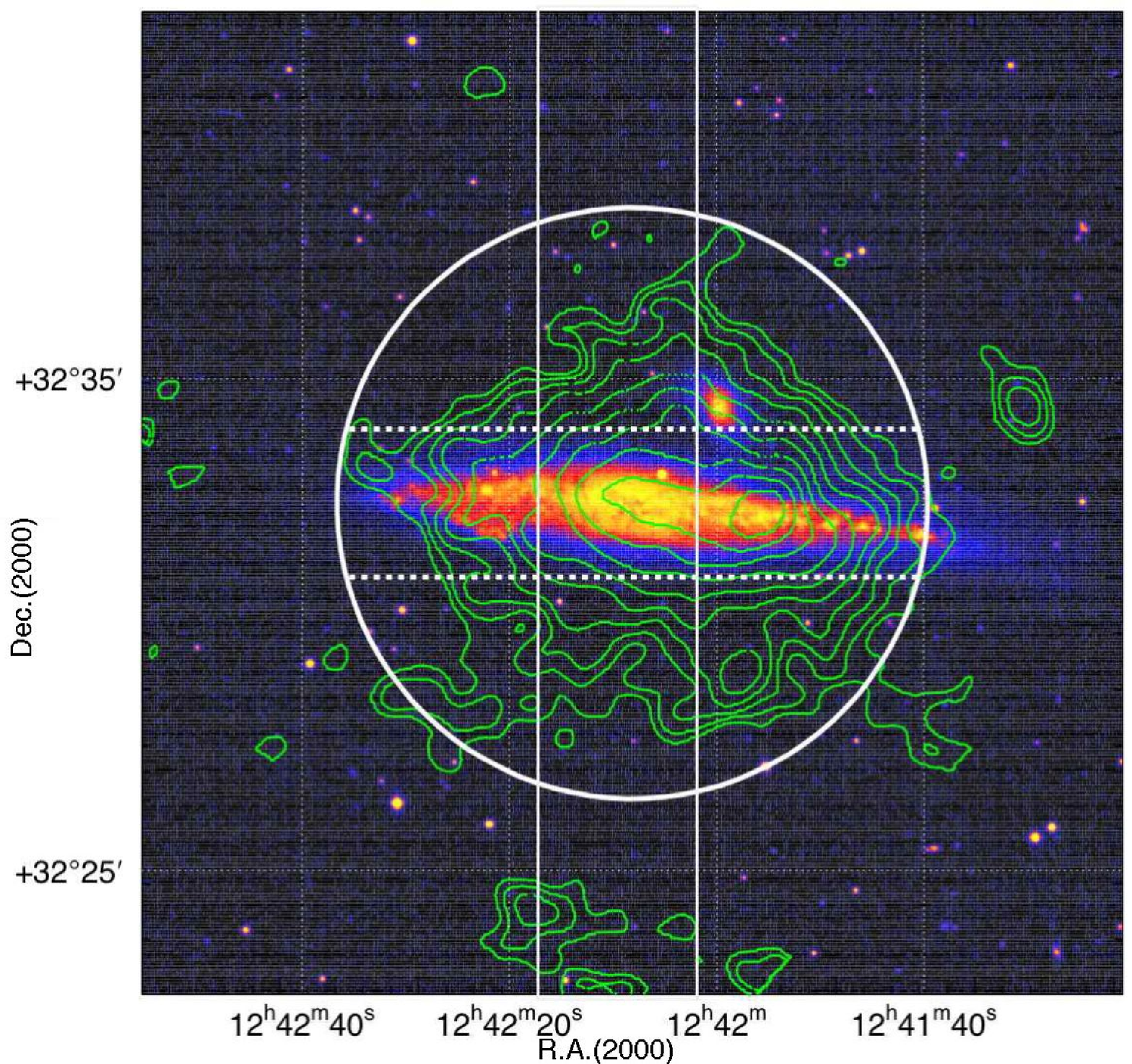}
\end{minipage}

\caption{
Left:The X-ray image of NGC 4631 observed with  Suzaku in the 0.6 -- 0.7 keV energy band. 
Right: The X-ray contour map in linear scale  from  0.5--2.0 keV overlaid on an optical image taken
by DSS.
For both images, the observed XIS0,1,3 images were added on the sky coordinate
after removing each calibration source region, and smoothed with
$\sigma=16$ pixel or $\simeq 17''$ Gaussian. Estimated components of
extragalactic X-ray background (CXB) and instrumental background (NXB)
were subtracted, and the exposure was corrected, though vignetting was
not corrected. The region where energy spectra were extracted are
shown by solid and dotted lines in the right figure, for the halo and disk regions,
respectively. Two Vertical  lines show the region where we took the surface
brightness profiles in section \ref{sec:HR}.}
\label{fig:img}
\end{figure*}

Suzaku carried out observations of the  starburst galaxy NGC~4631 in
November 2006 (PI: N. Y. Yamasaki) with an exposure of 81 ks. The
observation log is given in table~\ref{tab:log}.
The X-ray Imaging Spectrometer (XIS: \cite{koyama07}) instrument
consists of a back-illuminated (BI: XIS~1) CCD sensor and two
front-illuminated (FI: XIS~0, 3) sensors. The XIS was operated in the
Normal clocking mode (8~s exposure per frame), with the standard
$5\times 5$ or $3\times 3$ editing mode. The observed X-ray image in 
0.6 -- 0.7 keV band and XIS contour
image in 0.5--2.0~keV range overlaid
on the optical image by DSS is shown in figure~\ref{fig:img} .

\begin{table*}
\caption{Suzaku Observation logs for NGC~4631.}
\label{tab:log}
\begin{tabular}{lccccc} \hline 
Object & Seq. No. & Obs. date & \multicolumn{1}{c}{(RA, Dec)$^\ast$} &Exp.&After screening \\
&&&J2000& ks &(BI/FI) ks \\
\hline 
NGC~4631 & 801019010 & 2006-11-28 03:23 -- 2006-11-29  21:55& (\timeform{12h42m09.3s}, \timeform{+32D32'52''})& 81.1& 75.9/76.1\\
\hline\\[-1ex]
\multicolumn{6}{l}{\parbox{0.9\textwidth}{\footnotesize 
\footnotemark[$\ast$]
Average pointing direction of the XIS, written in the 
RA\_NOM and DEC\_NOM keywords of the event FITS files.}}\\
\end{tabular}
\end{table*}

\subsection{Data Reduction}

We used version 2.0.6 processing data \citep{mitsuda07}, and the
analysis was performed with HEAsoft version 6.4 and XSPEC 11.3.2aj.
In the analysis of XIS data, we selected ${\it ELEVATION} > 15^\circ$
of the data set to remove  stray-light from the day Earth limb, and the light
curve of each sensor in the 0.3--10~keV range with a 16~s time bin was
also examined so as to exclude periods of an anomalous event rate
greater or less than $\pm 3\sigma$ around the mean. After the above
screenings, the remaining exposures of the observations decreased by
$\sim7\%$ as shown in table~\ref{tab:log}. Event screening with
cut-off rigidity (COR) was not performed in our data.

In order to subtract the non X-ray background (NXB), we used a dark
Earth database, provided by the ``xisnxbgen'' ftools task
\citep{tawa08}. Although it is known that the optical blocking
filters of the XIS have gradually been contaminated by outgassing from
the satellite, we included these effects in the calculation of the
Ancillary Response File (ARF) by the ``xissimarfgen'' ftools task
\citep{ishisaki07}. We then generated two different ARFs for the
spectrum of each region, $A^{\makebox{\small\sc u}}$ and
$A^{\makebox{\small\sc b}}$, which respectively assumed uniform sky
emission and the observed XIS image. 
The task  wrote a value,  
"SOURCE\_REG\_RATIO",   which represents the  flux ratio 
in the assumed spatial distribution on the sky inside the spectral 
accumulation region to the entire model, in the ARF file header 
to evaluate the flux within the region.
Since the energy resolution has 
slowly degraded after the launch, due to radiation damage, this effect
was included in the Redistribution Matrix File (RMF) by the
``xisrmfgen'' ftools task. 

We reprocessed the HXD data with the CALDB files of 
2008-04-01 version, and applied data reduction by 
standard criteria.
The HXD data was cleaned by the Earth elevation angle $>5^{\circ}$,
the cut-off rigidity (COR) $>6$~GV, and was also processed by applying the
dead-time correction. After the screening, the exposure time of the
HXD data was 73.3 ks. We used the ae\_hxd\_pinxinome3\_20070914.rsp
file and ae801019010hxd\_pinnxb\_cl.evt as the response and the non
X-ray background (NXB) files, respectively, and also simulated the 100
Ms accumulation of the cosmic X-ray background data as described in
http://heasarc.gsfc.nasa.gov/docs/suzaku/analysis/pin\_cxb.html using
ae\_hxd\_pinflate3\_20070914.rsp as a response file.

\section{Analysis and Results}

\subsection{XIS spectra and the Galactic background}

We extracted spectra for three regions from the optical and X-ray images
as shown in figure~\ref{fig:img} :
\begin{enumerate}
\item Disk component: a rectangular region of $12'\times3'$ and $r<6'$
circle centered on (RA, Dec) = (\timeform{1h42m08s},
\timeform{+32D32'29''}) (35.5 arcmin$^{2}$).

\item Halo component: a circular region of $r<6'$ from the same center 
position of the disk component and outside the disk region (77.6 arcmin$^{2}$).

\item Background: outside of the above disk and halo region ($r>6'$).
\end{enumerate}

We first fitted the spectra of the background region and produced
the extra-galactic cosmic X-ray background (CXB) and Galactic
emission. This was because the O\emissiontype{VII} and
O\emissiontype{VIII} lines from the Galactic emission affected these
lines from NGC~4631 (see also \cite{sato07a}). We assumed either  one
or two temperature {\it apec} model for the Galactic emission
\citep{lumb02}, and tested the following two models: ${\it apec} +
phabs \times power\mathchar`-law$, and ${\it apec}_1 + phabs \times
({\it apec}_2 + power\mathchar`-law)$, where the {\it apec} models had
fixed metal abundance at 1 solar with zero  redshift and the absorption
column for "phabs" was fixed to the Galactic value. The 0.4--5.0 keV
spectra for  the BI and FI sensors for all regions were fitted
simultaneously, excluding an energy range of anomalous response around
the Si K-edge (1.825-1.840 keV)\@. The results of those fits are
given in table~\ref{tab:bgd}. In both cases, the surface brightness
of the CXB component are  consistent with the averaged value of $\sim
10$ photons cm$^{-2}$ s$^{-1}$ sr$^{-1}$ keV$^{-1}$ at 1 keV
\citep{gendreau95} with fluctuations for this sky area. Afterwards,
we adopted the two temperature model based on the F-test of these
results. The intensity of the O\emissiontype{VII} line is $4.4 \pm 0.8$
photons cm$^{-2}$s$^{-1}$sr$^{-1}$ (this unit is hereafter refereed to as the Line Unit or LU), 
which is almost the same as or
slightly smaller than the previously reported values (see
e.g.\ \cite{sato07b,mccammon02}).
If we subtract the energy spectrum of the background from those of the halo and disk regions, 
the resultant line intensities without correcting the absorption effect are 
$17.5^{+2.2}_{-4.1}$  LU for O\emissiontype{VII} and 
$12.0^{+1.9}_{-1.1}$  LU for O\emissiontype{VIII} in the halo and 
$5.1^{+1.0}_{-1.4}$  LU for O\emissiontype{VII} and 
$5.8^{+0.8}_{-0.7}$  LU for O\emissiontype{VIII} in the disk region.

\begin{table*}
\caption{ The best-fit parameters of the {\it apec} components
for the spectra in the background region of NGC~4631
with one or two temperature models (${\it apec}$) for Galactic emissions, 
and a {\it power-law model} for CXB. }
\label{tab:bgd}
\begin{tabular}{lcc}
\hline \hline
Parameters & ${\it apec_1} + {\it phabs}\times {\it power\mathchar`-law}$ &${\it apec_1} + {\it phabs}\times ({\it apec_2} + {\it power\mathchar`-law})$ \\
\hline
$kT_1$ (keV) & $0.145^{+0.008}_{-0.010}$& $0.114^{+0.010}_{-0.013}$ \\
${\it Norm}_1\,^\ast$ & $0.80\pm 0.13$ & $1.02\pm 0.42$ \\
$kT_2$ (keV) & -- &$0.310^{+0.096}_{-0.061}$ \\
${\it Norm}_2\,^\ast$ &-- & $0.13\pm 0.13$ \\
$\Gamma$ & $1.68^{+0.06}_{-0.06}$ &$1.58^{+0.07}_{-0.07}$ \\
${\it Norm}^\dagger$ & $0.93\pm0.05$ & $0.85\pm0.09$ \\
$\chi^2$/dof & 396/375 & 382/373 \\ \hline
\multicolumn{3}{l}{\parbox{0.94\textwidth}{\footnotesize 
\footnotemark[$*$] 
Normalization of the {\it apec} components
divided by the solid angle, $\Omega^{\makebox{\tiny\sc u}}$,
assumed in the uniform-sky ARF calculation (20$'$ radius),
${\it Norm} = \int n_{\rm e} n_{\rm H} dV \,/\,
(4\pi\, (1+z)^2 D_{\rm A}^{\,2}) \,/\, \Omega^{\makebox{\tiny\sc u}}$
$\times 10^{-20}$ cm$^{-5}$~arcmin$^{-2}$, 
where $D_{\rm A}$ is the angular distance to the source.}}\\
\multicolumn{3}{l}{\parbox{0.94\textwidth}{\footnotesize 
\footnotemark[$\dagger$] 
Normalization of the {\it power-law} component
divided by the solid angle same as the normalization of {\it apec},
in units of $10^{-8}~ {\Omega^{\makebox{\tiny\sc u}}}^-1$ photons keV$^{-1}$ cm$^{-2}$~s$^{-1}$~arcmin$^{-2}$ at 1 keV\@.}}\\
\end{tabular}
\end{table*}

In order to take into account both the existence of the Galactic component
itself and propagation of its statistical error, we simultaneously
fitted all (BGD, halo, disk) regions with the same model: $ {\it
phabs} \times {\it zphabs} \times ({\it vapec}_{\rm 1T or 2T} + {\it
zbremss} + {\it power\mathchar`-law} )+ {\it constant} \times({\it
apec}_1 + {\it phabs} \times {\it apec}_2 )$. In the model, {\it
phabs} corresponds to our Galactic absorption with fixed $N_{\rm H} =
1.27\times10^{20}$cm$^{-2}$. The term $({\it apec}_1 + {\it phabs}
\times {\it apec}_2 )$ represents the Galactic component with a
normalization factor to keep a  uniform surface brightness. Based on
the previous fits, we fixed the temperatures of the Galactic emission,
${\it apec}_1$ and ${\it apec}_2$, to be 0.114 keV and 0.310 keV,
respectively, with 1 solar abundance and zero redshift.
The ${\it power\mathchar`-law}$  component which represents the CXB
was  set to be common for all three regions.

For the ISM emission from NGC 4631, {\it zphabs} corresponds to the
intrinsic absorption. The hot X-ray halo around NGC 4631 is modeled
either with a one or two temperature collisionally ionized plasma with
various metal abundances given by the {\it vapec} model. Note that the
abundances were common for both temperature components, and we
combined the metal abundances into five groups as  O, Ne, (Mg \& Al),
(Si, S, Ar, \& Ca), and (Fe \& Ni). As we found that the continuum
emission above 2 keV cannot be fitted only with the CXB, we added a {\it
zbremss} model of $kT= 10$ keV for the low-mass X-ray binary (LMXB)
component. For the spectrum of the background region, {\it zphabs}
and the normalizations of {\it vapec} and {\it zbremss} were all fixed to
be 0\@.

\begin{figure*}
\begin{minipage}{0.33\textwidth}
%%\FigureFile(\textwidth,\textwidth){../sato0319/bf_all_v1s_disc.eps}
\FigureFile(\textwidth,\textwidth){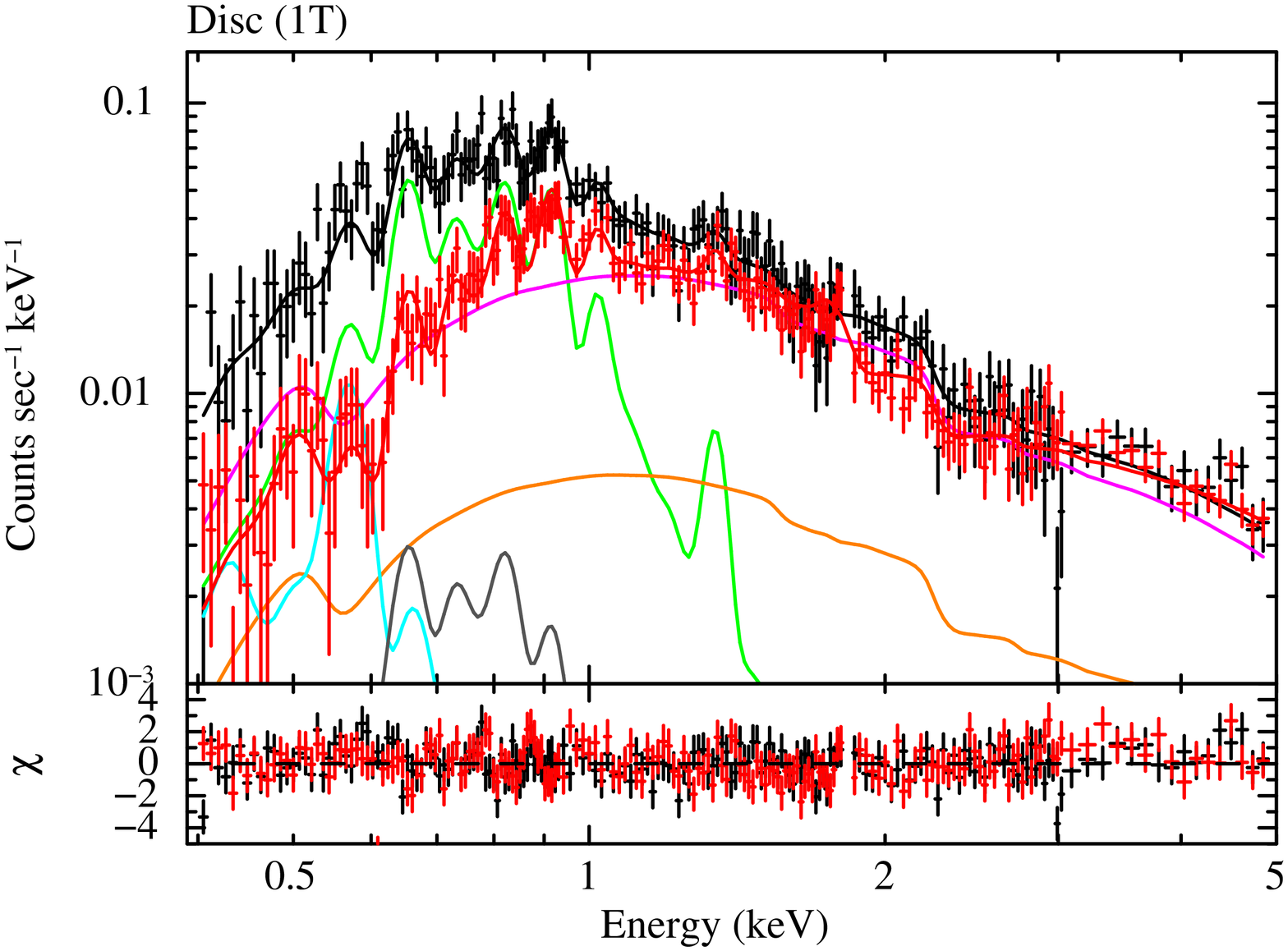}
\end{minipage}\hfill
\begin{minipage}{0.33\textwidth}
%%\FigureFile(\textwidth,\textwidth){../sato0319/bf_all_v1s_halo.eps}
\FigureFile(\textwidth,\textwidth){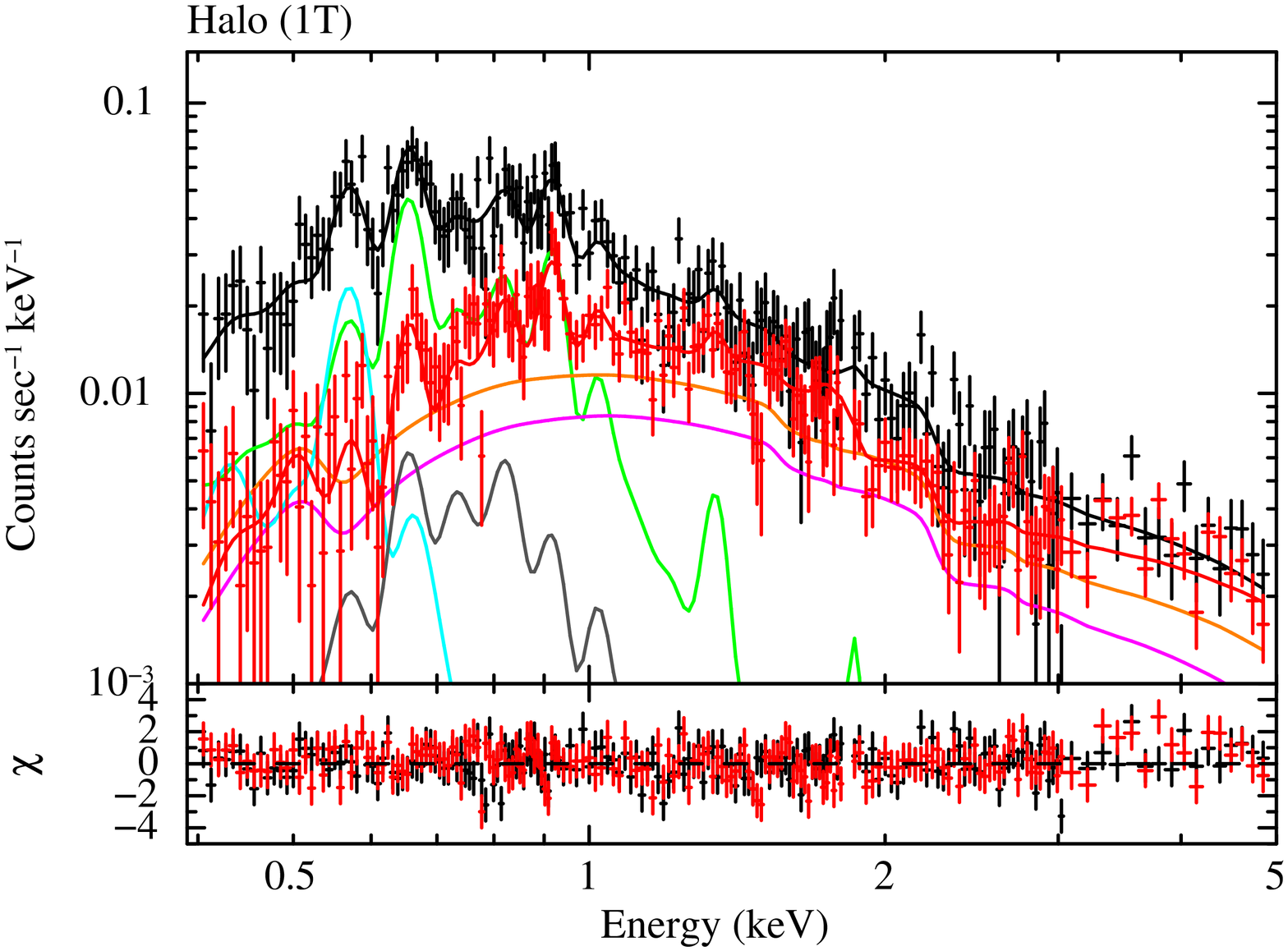}
\end{minipage}\hfill
\begin{minipage}{0.33\textwidth}
%%\FigureFile(\textwidth,\textwidth){../sato0319/bf_all_v1s_bgd.eps}
\FigureFile(\textwidth,\textwidth){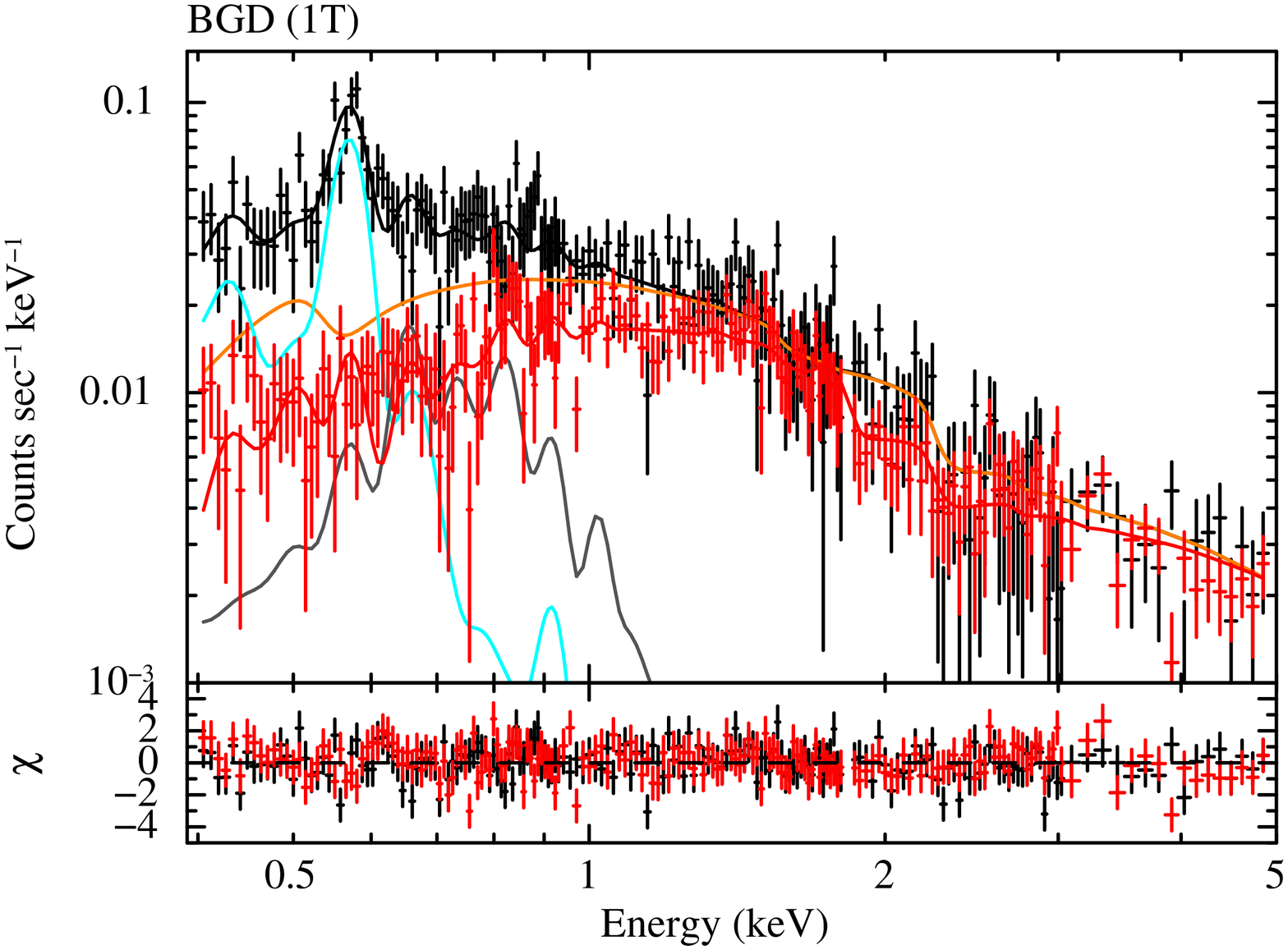}
\end{minipage}

\begin{minipage}{0.33\textwidth}
%%\FigureFile(\textwidth,\textwidth){../sato0319/bf_all_v6s_disc.eps}
\FigureFile(\textwidth,\textwidth){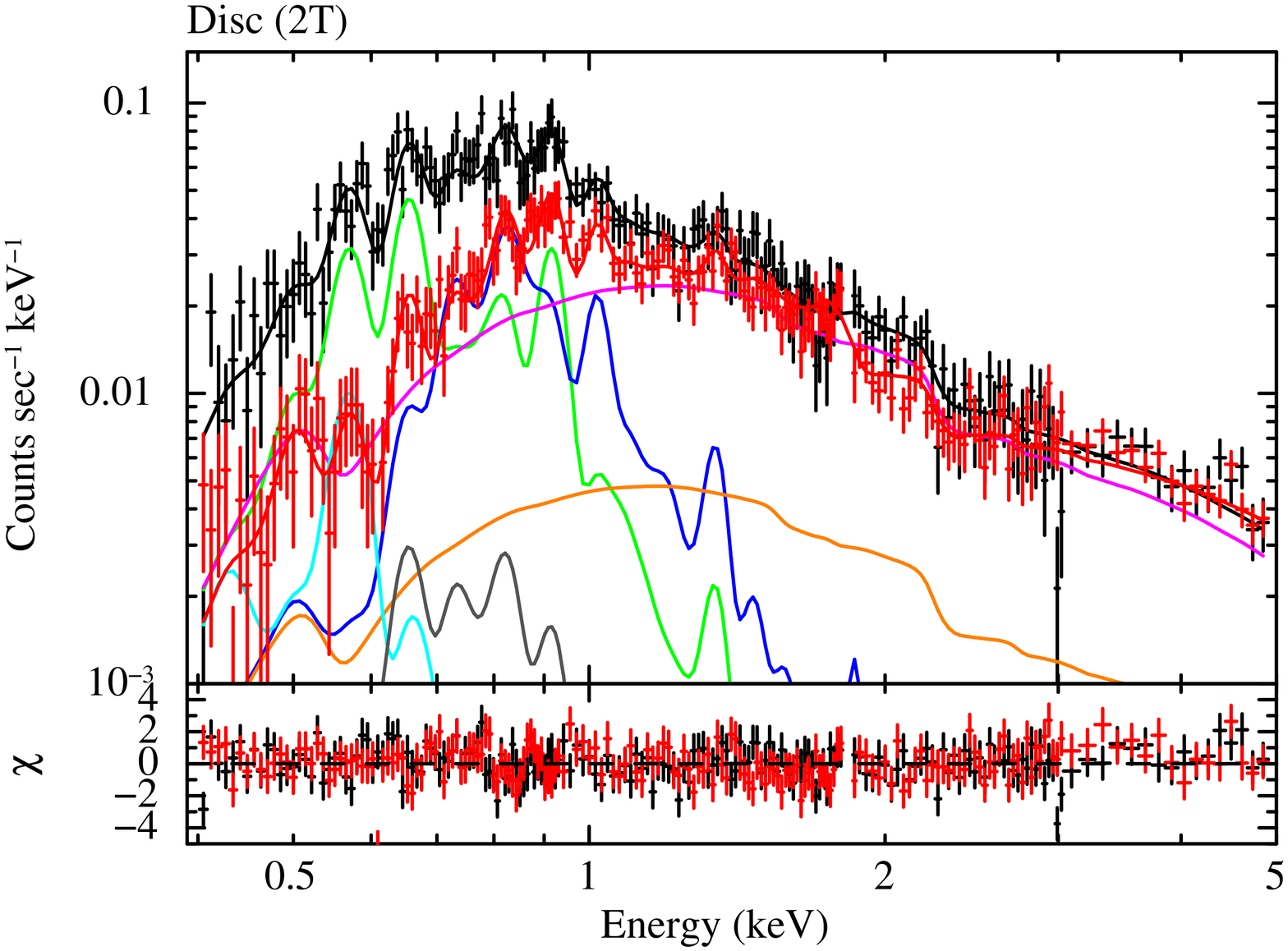}
\end{minipage}\hfill
\begin{minipage}{0.33\textwidth}
%%\FigureFile(\textwidth,\textwidth){../sato0319/bf_all_v6s_halo.eps}
\FigureFile(\textwidth,\textwidth){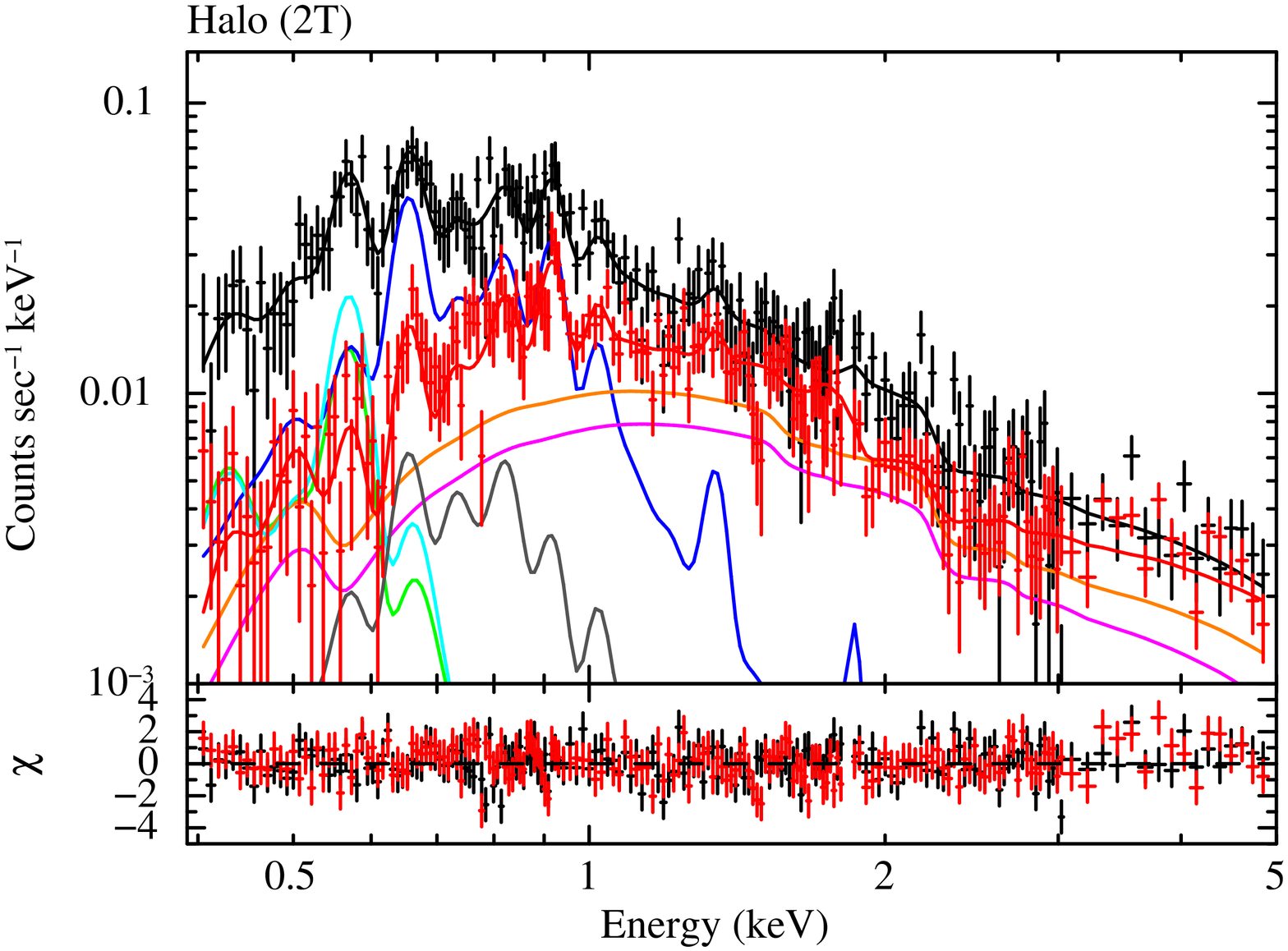}
\end{minipage}\hfill
\begin{minipage}{0.33\textwidth}
%%\FigureFile(\textwidth,\textwidth){../sato0319/bf_all_v6s_bgd.eps}
\FigureFile(\textwidth,\textwidth){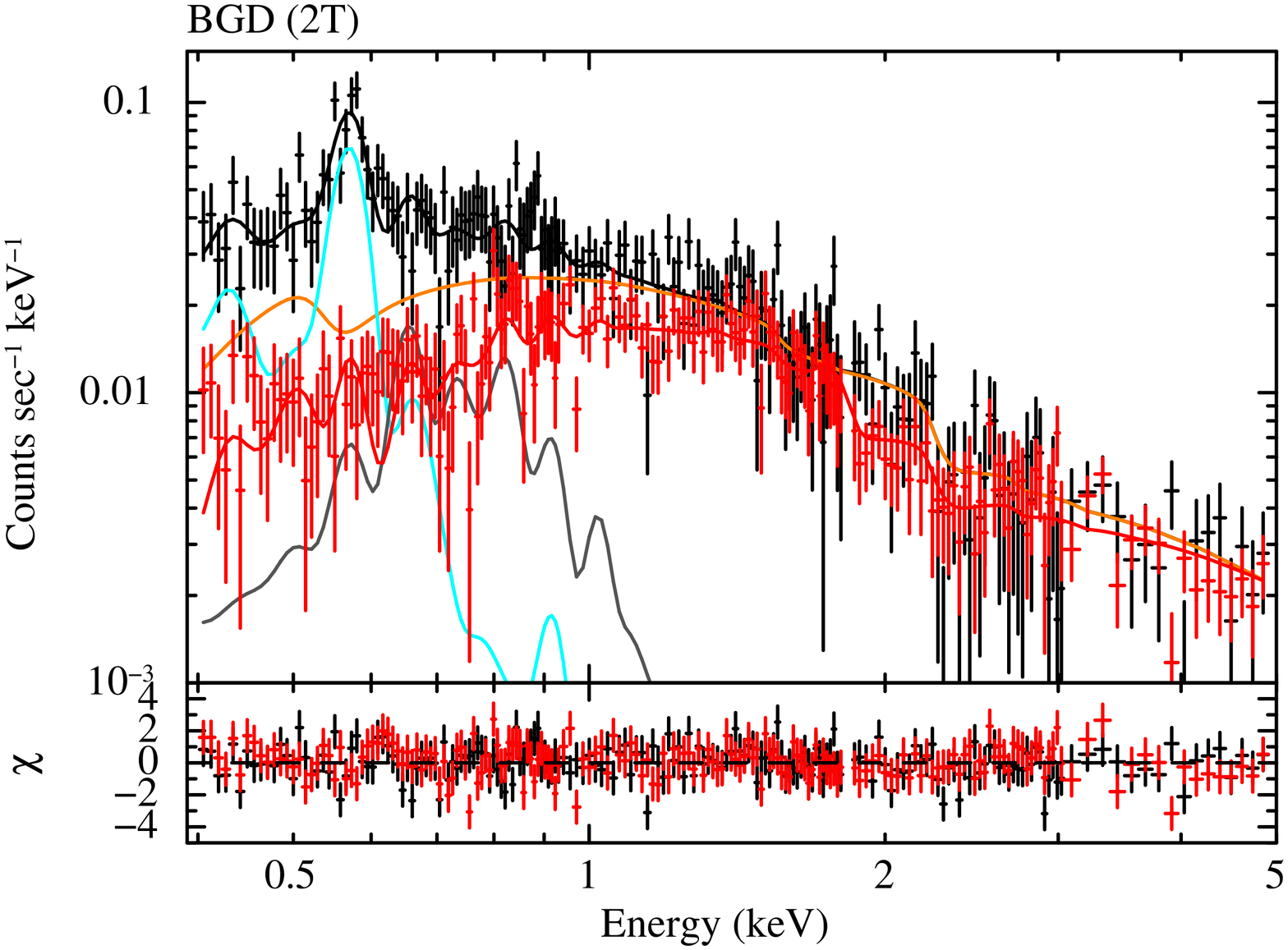}
\end{minipage}

\caption{The panels show the observed spectra 
after subtracting the NXB component for all regions of
NGC~4631 which are denoted in the panels, and they are plotted by black and 
red  crosses for BI and FI, respectively. 
The spectra not corrected for the integrated area.
They are fitted with
the model: $ {\it phabs} \times {\it zphabs} \times ({\it vapec}_{\rm
1T or 2T} + {\it zbremss} + {\it power\mathchar`-law} )+ {\it
constant} \times({\it apec}_1 + {\it phabs} \times {\it apec}_2 )$ as
shown by black and red lines for the BI and FI spectra. For
simplicity, only the model components for BI spectra are shown. Green
and blue lines are the ISM component by {\it vapec}, cyan and grey are
the Galactic background emission by ${\it apec}_1$ and ${\it apec}_2$,
magenta and orange are the LMXB and CXB component, respectively. The
lower panels show the fit residuals in unit of $\sigma$.
}\label{fig:fit_all}
\end{figure*}

The resultant spectra are shown in figure \ref{fig:fit_all} and
parameters are summarized in table \ref{tab:result}.  For the single temperature
ISM model, the resultant temperatures are $kT= 0.297$ keV for the disk
and 0.290 keV for the halo, respectively. The emission line profiles
for Si and S are difficult to  distinguish, and the abundance value is
almost consistent with 0. The fit statistics shown in table
\ref{tab:result} support the two temperature model. 
The two temperature model gives larger absorption column and 
larger intrinsic luminosity than the single temperature model. 
The power-law
index of the CXB component was obtained to be $1.48^{+0.06}_{-0.03}$
and the intensity of the LMXB component for the disk corresponds to
$\sim 3\times10^{39}$ erg s$^{-1}$, which is consistent with the 
$L_{X}-M$ relation for spiral galaxies \citep{gilfanov04}. A fraction
of the LMXB component in the halo region is considered to be due to
the extended tail of the telescope PSF\@.
 We consider that
the two temperature model actually approximates multi-temperature
emission, and it is represented by emission peak temperatures for
O\emissiontype{VII} and O\emissiontype{VIII} with lower metal
abundance. However, we do not further carry on detailed
multi-temperature analysis given  the statistical quality of the
data.

We note   possible  systematic errors due to the assumption of the 
contamination model of the  XIS. When we change the thickness of the 
contamination by $\pm$10 \%,  the flux and the temperature 
of the $kT_{1}$ component of the halo changes 
about 30\%, and 3\% respectively. Thus, the abundances of metals also 
change by $\sim 10$ \%. The $kT_{2}$ of the halo and $kT_{1}$ and $kT_{2}$ of the disk
change only by  1\%. 
 
We briefly compare with the previous results for NGC 4631
here. \citet{wang01} reported higher temperatures ($0.18\pm 0.02$ and
$0.61\pm 0.1$ keV) and lower ($0.08\pm0.04$) metal abundances after
subtracting the blank field of the similar foreground absorption.
\citet{tuellmann06} observed with XMM-Newton and detected halo
emission up to 9.1 kpc from the disk. They sliced the halo into 8
regions and fitted the spectra with 2-temperature Raymond-Smith plasma
model with the cosmic metal abundance. The resultant temperatures are
$kT_{soft} \sim 0.17-0.07 $ keV and $kT_{hard}\sim 0.28-0.2$ keV\@.
Considering the different plasma code and assumed metal abundance,
the agreement of the results is fairly good.

\begin{table*}
\caption{Summary of the best-fit parameters of one or two {\it vapec}
components by simultaneous fit of all regions. }\label{tab:result}
\begin{center}
\begin{tabular}{lcccc}
\hline \hline
& \multicolumn{2}{c}{ISM 1T} & \multicolumn{2}{c}{ISM 2T} \\
& Disk & Halo & Disk & Halo \\
\hline
$kT_1$ (keV) &     $0.297^{+0.009}_{-0.016}$&$0.290^{+0.027}_{-0.053}$&$0.209^{+0.025}_{-0.016}$ &$0.096^{+0.065}_{-0.015}$\\
{\it Norm}$_1^\ast$ & $4.5^{+2.4}_{-2.2}$&$0.9^{+1.2}_{-0.8}$ &$8.0^{+2.3}_{-4.5}$ & $4.5^{+13.4}_{-4.2}$ \\
flux$_{1}^\dagger$ (erg cm$^{-2}$~s$^{-1}$) &$2.2\times10^{-13}$ &$1.3\times10^{-13}$ & $1.6\times10^{-13}$ & $2.3\times10^{-14}$ \\
$kT_2                     $ (keV) & -- & -- &$0.480^{+0.073}_{-0.101}$ &$0.300^{+0.016}_{-0.027}$ \\
{\it Norm}$_2^\ast$ & -- & -- & $2.3^{+1.7}_{-1.3}$ &$2.2^{+2.2}_{-1.3}$ \\
flux$_{2}^\dagger$  (erg cm$^{-2}$~s$^{-1}$)& -- & -- & $1.1\times10^{-13}$ & $1.4\times10^{-13}$\\
O (solar) & $0.97^{+0.81}_{-0.55}$ &$1.34^{+1.06}_{-0.91}$ &$0.73^{+0.69}_{-0.38}$ &$0.81^{+0.55}_{-0.43}$ \\
Ne (solar) &$1.92^{+1.70}_{-1.25}$ &$2.18^{+5.34}_{-1.37}$ &$1.60^{+2.11}_{-0.72}$ &$1.09^{+1.14}_{-0.55}$\\
Mg, Al (solar) &$1.85^{+3.41}_{-1.01}$ &$2.29^{+14.06}_{-1.79}$ &$1.26^{+0.70}_{-0.65}$ &$0.98^{+1.14}_{-0.69}$ \\
Si, S, Ar, Ca (solar) &$0.45^{+2.55}_{-0.45}$ &$5.98^{+20.57}_{-4.80}$ &$0.60^{+2.65}_{-0.60}$ & $2.17^{+0.83}_{-2.17}$ \\
Fe, Ni(solar) & $1.09^{+0.75}_{-0.64}$ & $0.85^{+1.47}_{-0.63}$ &$0.93^{+0.64}_{-0.41}$ & $0.46^{+0.31}_{-0.23}$\\
extra absorption (cm$^{-2}$)&$3.4^{+4.0}_{-2.0}\times10^{20}$ &$0.0^{+4.6}_{-0.0}\times10^{20}$ &$8.6^{+4.4}_{-4.0}\times10^{20}$&$6.4^{+4.4}_{-5.7}\times10^{20}$  \\
LMXB flux $^\dagger $ (erg cm$^{-2}$~s$^{-1}$) &$3.0\times10^{-13}$  &$8.5\times10^{-14}$  & $2.6\times10^{-13}$ & $7.8\times10^{-14} $\\
Hot gas luminosity$^\ddagger$ (erg s$^{-1}$)& $1.9 \times10^{39}$ &$0.9\times10^{39}$ &$3.1\times10^{39}$ &$1.6\times10^{39}$ \\
$\chi^2$/dof & \multicolumn{2}{c}{1198/1114} & \multicolumn{2}{c}{1172/1109} \\ \hline
\multicolumn{5}{l}{\parbox{0.8\textwidth}{\footnotesize 
\footnotemark[$*$] 
Normalization of the {\it vapec} component scaled with a factor of
{\sc source\_ratio\_reg} / {\sc area}, which is ${\it
Norm}=\frac{\makebox{\sc source\_ratio\_reg}}{\makebox{\sc area}} \int
n_{\rm e} n_{\rm H} dV \,/\, [4\pi\, (1+z)^2 D_{\rm A}^{\,2}]$ $\times
10^{-20}$~cm$^{-5}$~arcmin$^{-2}$, where $D_{\rm A}$ is the angular
distance to the source.\\
\footnotemark[$\dagger$] 
Flux within the accumulated region  between 0.5 and 2 keV \\
\footnotemark[$\ddagger$] 
Intrinsic luminosity   between 0.5 and 2 keV 
}}\\
\end{tabular}
\end{center}

\end{table*}

\subsection{Hardness ratio} \label{sec:HR}

\begin{figure*}
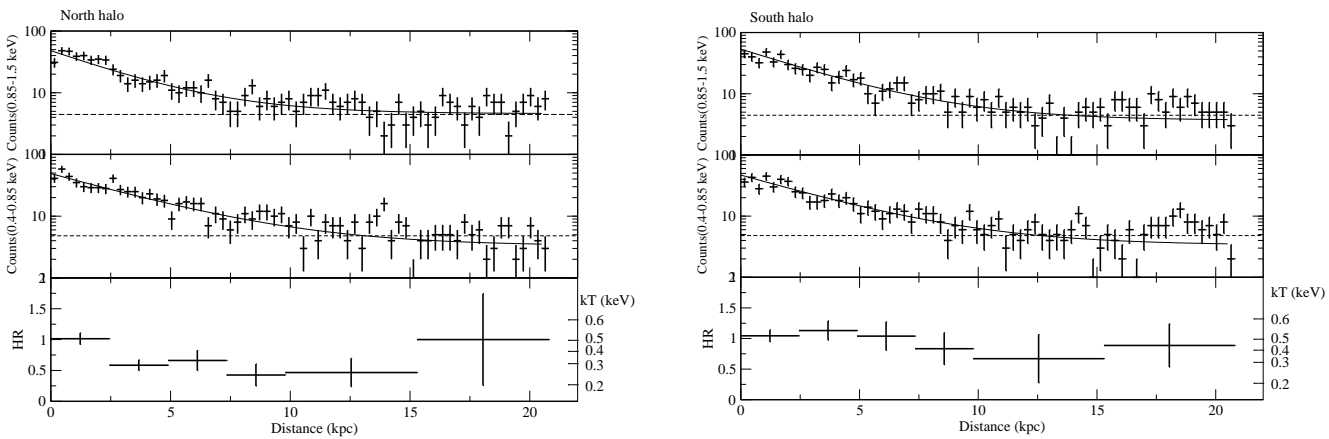

\vspace{5mm}
\begin{minipage}{0.49\textwidth}
%%\FigureFile(0.95\textwidth,\textwidth){North.eps}
\FigureFile(0.95\textwidth,\textwidth){figure3a.eps}
\end{minipage}
\hfill
\begin{minipage}{0.49\textwidth}
%%\FigureFile(0.95\textwidth,\textwidth){South.eps}
\FigureFile(0.95\textwidth,\textwidth){figure3b.eps}
\end{minipage}
\caption{Count rate profile of hard (0.85 -- 1.5 keV) and
soft (0.4 -- 0.85 keV) X-ray bands without subtracting background 
and the hardness ratio in the north (left) 
and the south (right) side. The region is shown in figure \ref{fig:img}.
Dashed lines in the count rate profile are drawn to show the averaged rate taken from the 
background region for the spectral fitting.}
\label{fig:HR}
\end{figure*}

We produced count rate profiles in a rectangular region of $7.5 \times
42$ kpc$^{2}$ as shown in figure \ref{fig:img} in two X-ray energy
bands, 0.4 -- 0.85 and 0.85 -- 1.5 keV. With Chandra,
\citet{strickland04} also showed surface brightness profile in 0.3 --
1 keV along the minor axis and found that an exponential model was
preferred over the  Gaussian or power-law models. We fitted the profile
with an exponential + constant model as shown in figure \ref{fig:HR},
in which the constant component represents the sum of the X-ray and non
X-ray backgrounds. The fit well represent the data, and can not be
rejected within a confidence level of 95\%. The constant levels are
consistent within errors to the same for the north and south sides.

The obtained scale heights are summarized in table
\ref{tab:scaleheight}. These values are larger than those by
\citet{strickland04} and could be due to the difference of the data
region used in the fit. The scale height values in table
\ref{tab:scaleheight} are consistent with each other to within 90\%
errors, but those in the hard band tend to be smaller than those in
the soft band for both sides of the disk. This corresponds to a
spectral softening, consistent with the feature obtained by the
spectral fit with XMM-Newton \citep{tuellmann06}.

 We calculated the hardness ratio for the exponential component after
subtracting the constant levels with errors, and the results are
plotted in figure \ref{fig:HR}. If we assume a single temperature
thermal emission for the disk with abundances shown in table
\ref{tab:result}, temperatures of $kT=0.2, 0.3,$ and $ 0.4 $ keV give
HR values of 0.268, 0.610, and 0.809, respectively. Toward the outer
region at $>10$ kpc from the disk, $kT$ is  consistent with $>0.2$
keV\@. The small or no  decline of the temperature may imply that the gas is
adiabatically expanding into a vacuum with small mechanical work.

\begin{table*}
\caption{Scale height of the X-ray emission}
\label{tab:scaleheight}
\begin{center}
\begin{tabular}{llcc}
\hline
\multicolumn{2}{l}{Scale height (kpc)} & North side & South side \\ 
\hline
Hard band&(0.85 -- 1.5 keV)& 3.01 $\pm$ 0.47 & 3.33$\pm$ 0.47 \\
&($\chi^2/dof$) & (66.02/65) & (81.30/65) \\ 
Soft band&(0.40 -- 0.85 keV)& 3.85 $\pm$0.58 & 3.72 $\pm$ 0.59 \\
&($\chi^2/dof$) &(79.24/65) & (87.08/65) \\
\hline 
\end{tabular}
\end{center}
\end{table*}

\subsection{Search for the hard X-ray emission}

Since NGC 4631 is accompanied by a strong radio halo which is likely
to be due to  synchrotron emission by relativistic electrons, one can expect
inverse Compton emission in the hard X-ray band. We searched for hard
X-ray emission with the PIN detector of the HXD over the $34'$ fov. After
subtraction of the NXB component estimated by the standard modeling
method in the current processing, the energy spectrum was consistent
with the expected CXB level. Due to the current uncertainty of the
NXB estimation, we scaled the normalization of the NXB by $-3\%$
\citep{kokubun07}, and the spectrum was fitted by ${\it phabs} \times {\it
power\mathchar`-law}$ model with a fixed photon index to be 2.0 in 12--60 keV\@.

The resultant upper limit corresponding to $1 \sigma$ level of the
flux was $5.5\times10^{-12}$ erg cm$^{-2}$ s$^{-1}$ in 12--60 keV, 
with the estimated contribution from LMXBs with an energy spectrum of 
$kT= 10$ keV is $4\times10^{-13}$ erg cm$^{-2}$ s$^{-1}$.
The Very Large Array (VLA) observations at 1.49 GHz indicated a radio
halo of NGC~4631 with a brightness of about 1.22 Jy \citep{hummel90}.
If we assume an inverse Compton process with 2.7 K photons scattered
by the same relativistic electrons responsible for the radio halo, the
strength of the magnetic field is constrained. Following the
prescription given by \citet{harris79}, we derived the lower limit of
the magnetic field strength to be $B>0.5~{\rm \mu G}$, which is
consistent with the previous estimation of 5 $\mu$G from the  polarized
component of the radio emission \citep{hummel91}.

\section{Discussion}

\begin{figure}
%%\FigureFile(\columnwidth,\columnwidth){Z_O_bw.eps}
\FigureFile(\columnwidth,\columnwidth){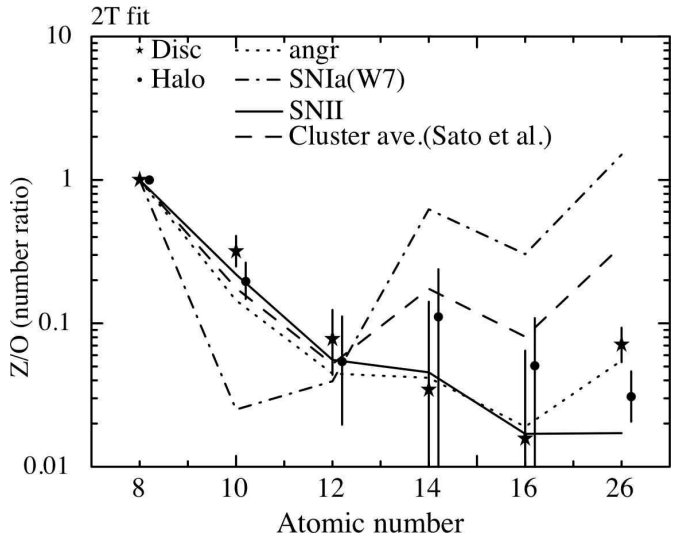}
\caption{Number ratios of Ne, Mg, Si, S, and Fe to O for disk and halo
regions.  Solid, dotted, dashed, and dot-dashed lines correspond to the
number ratios of metals to O for  abundance patterns  of
SNe II yield of \citet{nomoto06}, solar abundance by \citet{anders89}, 
cluster average in \citet{sato07a},
and  SNe Ia yield of \citet{iwamoto99},
respectively.}
\label{fig:ratio}
\end{figure}

 Suzaku observations of NGC 4631 showed significant emission lines
from an extended halo region, including those from O for the first
time. We examined an abundance pattern for the X-ray halo in the form
of the number ratio to O as shown in figure \ref{fig:ratio}. 
To make this plot, Z/O ratios are determined by two parameter errors 
from the simultaneous fit of all regions as described in section  3.1. 
In this
plot, the SN Ia yields were taken from W7 model in \citet{iwamoto99}.
For SN II, \citet{nomoto06} gave an average yields for the Salpeter's
IMF of stellar masses from 10  to 50 $M_{\odot}$ with the progenitor
metallicity of $Z=0.02$. If the  metallicity of progenitors increases to 
$Z=1$, relative abundances of ejecta increase by at most 20\% for Fe/O.
Also the solar abundance template of 
\citet{anders89}, and the average abundance pattern for 4 clusters and
groups \citep{sato07a} are plotted together. The pattern in clusters
is well fitted by a combination of SN Ia and SN II with a ratio of
1:3.5. The metal abundance of NGC 4631 halo disagrees with that of SN
Ia and the cluster average, but is  consistent with the SN II yields.
In the disk component, Fe shows relatively higher abundance and the
abundance pattern is consistent with the Solar abundance given by
\citet{anders89}.

For the study of the energetics of the gas, we estimate the density,
mass, and total energy of the X-ray emitting gas. As shown in figure
\ref{fig:HR}, the X-ray profile of the halo is very smooth, and there
is no boundary recognized between the disk and the halo. Hereafter,
we will not discriminate the disk and the halo component and treat
them at the same time.
We assume a simple exponential model of $n(z)=n_{0}\times \exp(-z/h)$
with $z$ indicating the distance from the galactic plane, within a
radius of 10 kpc for the coronal gas. Since the scale height of the
surface brightness is about 3.5 kpc as shown in Table
\ref{tab:scaleheight}, we can adopt the density scale height to be $h=
7$ kpc. A single temperature model gave us a lower limit of the
density at the disk as $n_{0}= 2\times10^{-3}$ cm$^{-3}$, and the
pressure as $n_{0}T=7\times10^{3}$ cm$^{-3}$ K, which are consistent
with the previous values by \citet{wang95,wang01}. With this density,
the total mass of the X-ray emitting gas is $1.3\times10^{8}
M_{\odot}$, and the stored thermal energy is $2\times10^{56}$ erg.
Assuming an O abundance of 0.8 solar, the O mass in the hot gas in
both the disk and the halo is $\sim 10^{6} M_{\odot}$.

Since the cooling time in the disk is about $\sim 6\times 10^{8}$ yr
assuming the cooling function of  \citet{sutherland93}, the required
energy input rate is $3\times10^{47}$ erg yr$^{-1}$ and a mass transfer
rate of $\sim 0.2 M_{\odot}$ yr$^{-1}$, respectively. If one employs
a flow time to a radius of 10 kpc, it is about $5 \times 10^7$ yr and
10 times higher rates for energy input and mass transfer are implied.
\citet{wang95} estimated the maximum mass flow rate from the density
multiplied by the sound velocity to be $1.4 M_{\odot}$ yr$^{-1}$.
Based on the UV observation of O\emissiontype{VI} line, \citet{otte03}
estimated the flow rate to be $0.48 \sim 2.8 M_{\odot}$ yr$^{-1}$
assuming a cooling flow model by \citet{edgar86}. Considering the
differences in the assumed physical process and condition,  mass flow
rates are in good agreement around an approximate value of $1
M_{\odot}$ yr$^{-1}$.

The most likely source of the energy and the material in the
coronal gas is SNe.
The SFR is estimated by the FIR luminosity as $3 M_{\odot}$ yr$^{-1}$
\citep{strickland04}. As NGC ~4631 has an edge-on morphology, the
emission could be underestimated due to absorption through the disk.
\citet{persic04} proposed another method to estimate the SFR using the
X-ray luminosity of high mass X-ray binaries (HMXB) in the 2--10 keV
band. A typical luminosity ratio of HMXB/LMXB of 0.2 gives a SFR of $1.2
M_{\odot}$ yr$^{-1}$. If we assume that all the flux above 2 keV comes
from HMXBs, it gives an upper limit as $6 M_{\odot}$ yr$^{-1}$. We
note that a spectral fit with a power-law component with $\Gamma=1.2$
as suggested by \citet{persic04} requires a steep $\Gamma =1.55$ CXB 
component. 
The two independent
estimates  give a consistent SFR of $\sim 3 M_{\odot}$ yr$^{-1}$,
which is almost the same as or a little  less than the level in our
Galaxy, i.e.\ $\sim 5 M_{\odot}$ yr$^{-1}$ \citep{zinnecker07}.
Assuming the Salpeter's IMF integrated over 0.1 to $50 M_{\odot}$, a
SFR of $3 M_{\odot}$ yr$^{-1}$ gives a SN II rate of $7\times10^{-3}$
SN yr$^{-1}$. In addition to SN II, SN Ia of $2.5 \times10^{-3}$ SN
yr$^{-1}$ is expected for the total mass of $2.6\times10^{10}
M_{\odot}$ and the SFR of $3 M_{\odot}$ yr$^{-1}$ \citep{sullivan06}.

If SNe are the source of the energy and the mass of the halo gas
estimated above, one SN needs to supply an energy of $3\times10^{49}$
erg 
and a mass of $\sim 20 M_{\odot}$ including an  O mass of $0.2 M_{\odot}$.
Therefore, 3\% of the typical explosion energy of $10^{51}$erg and $20
M_{\odot}$ from the ejecta and ambient material have to escape into
the halo. These values indicate that the halo gas is produced very
efficiently. Since one SN produces 0.14 and $1.8 M_{\odot}$ of O for
SN Ia and SN II, respectively \citep{iwamoto99}, a supply of $0.2
M_{\odot}$ of O seems plausible. The abundance pattern in the halo
is, however, well represented by the SN II products, although the
above estimated rate between SN Ia and SN II is $\sim 1:3$. It may
imply the selective escape of SN II ejecta into the halo gas, which can 
result  if the occurrence of SN II is concentrated in the starforming
region. In this case, superbubbles are formed and the metal-rich hot
gas escapes along a chimney to the halo space \citep{norman89}.

We must note two possibilities which may bias the abundance and
temperature estimate. \citet{lallement04} pointed out the possible
contribution of charge exchange (CX) processes  between the galactic wind
and gas clouds in the halo. The CX spectrum is dominated by emission
lines, and it tends to decrease the apparent temperature of the halo,
$kT_{1}$ in table \ref{tab:result}. The process was seriously
evaluated for the M82 halo \citep{ranalli08} and M82 ``cap'' region
\citep{tsuru07}, and they found that almost all of the
O\emissiontype{VII} triplet could be produced by CX\@. In the case of
NGC 4631, a dust arch with a mass of a few time $10^{8} M_{\odot}$ was
discovered in the halo, but there seemed to be no connection between the X-ray
emission and the dust arch \citep{taylor03}. Thus, the contribution
from CX might not be so large, considering the low velocity of the
outflow and the low density of the neutral material, which is also
suggested by the slow decline of the X-ray temperature. We
must  wait for improved spectroscopic observation to distinguish the
physical process of the emission. 

Another possibility is the role of dust. Dust of silicates
(Mg$_{2}$SiO$_{4}$) and other forms with a mass of $\sim 5 M_{\odot}$
SN$^{-1}$ can be formed 300--600 days after SN II explosions
\citep{todini01}. Since it requires more than $10^{7}$ yr to
evaporate in the low-density environment with $n=10^{-3}$ cm$^{-3}$
\citep{tielens94}, a significant amount of metals may be held in
dust. We hope that future high resolution X-ray spectroscopy will be
able to show the ionization condition of plasmas in the galactic winds
more precisely.

\section{Conclusion}
We determined the temperature and metal abundance of the X-ray
emitting halo gas around NGC 4631. The total energy, mass, and
metals in the halo can by supplied by SNe with the currently
estimated SFR, if the outflow efficiently carries metal-rich gas from
the starforming regions into the halo. The effect of neutral material
and dust should be taken into account to understand the plasma
properties in the halo.

\section*{Acknowledgement}
The authors acknowledge  Dr. Tai Oshima for his information 
about detailed Chandra data analysis and support for the observation.
Part of this work was financially supported by the Ministry of
Education, Culture, Sports, Science and Technology of Japan,
Grant-in-Aid for Scientific Research No. 20340041, 20340068 and 19840043.

The Digitized Sky Survey was produced at the Space Telescope Science
Institute under U.S. Government grant NAG W-2166. This research has
made use of the NASA/IPAC Extragalactic Database (NED) which is
operated by the JPL, under contract with NASA.
%%%
% See the manual for the detail.
%%%

\end{document}